# Closing the Gap: Can Novice Statistics and Data Science Students Collaborate as Effectively as an Expert?

Jessica L. Alzen
Center for Assessment, Design, Research and Evaluation, University of Colorado Boulder,
2445 Kittredge Loop Dr., Boulder, CO 80305,
Jessica.Alzen@colorado.edu, ORCiD: 0000-0002-1706-2975

Ilana M. Trumble
Department of Information Science, University of Colorado Boulder,
1045 18th Street, Boulder, CO 80309,
Ilana.Trumble@colorado.edu, ORCiD: 0000-0003-0531-6584

Kimberly J. Cho
Center for Assessment, Design, Research and Evaluation, University of Colorado Boulder,
2445 Kittredge Loop Dr., Boulder, CO 80305,
Kimberly.Cho@colorado.edu

Eric A. Vance (Corresponding Author)
Laboratory for Interdisciplinary Statistical Analysis, Department of Applied Mathematics,
University of Colorado Boulder,
1111 Engineering Dr., Boulder, CO 80309,
Eric.Vance@colorado.edu, ORCiD: 0000-0001-5545-1878

# Closing the Gap: Can Novice Statistics and Data Science Students Collaborate as Effectively as an Expert?

## Abstract

The ASCCR (Attitude-Structure-Content-Communication-Relationship) framework was recently developed to teach collaboration skills to statisticians and data scientists. However, its effectiveness in real-world settings has not yet been systematically evaluated. To assess this, we evaluated novice undergraduate and graduate students' performances in initial collaboration meetings with real domain experts and compared them to an expert collaborator. Using video recordings, rubric scores, and domain expert feedback surveys, we found that novices performed surprisingly well compared to the expert. Specifically, novices scored nearly as well as the expert on the Attitude, Structure, and Relationship components of the ASCCR framework. Although novices did not initially perform as well on the Content or Communication aspects, they were able to close the gap. By the end of the collaboration projects, the novices had higher overall domain expert feedback scores than the expert. The primary implication of our study is that novices can become effective collaborators in a very short time. We discuss our findings' practical implications and provide recommendations for integrating the ASCCR framework into statistics and data science collaboration, consulting, and capstone courses.

# 1. Introduction

Interdisciplinary research is necessary for innovation, exploration, and investigation of difficult questions across fields (Freeth and Caniglia 2020; National Academies of Scienes Engineering and Medicine 2004). Despite the fact that interdisciplinary collaboration is necessary and inescapable in many contexts, researchers are often underprepared to engage successfully in interdisciplinary collaboration work (Freeth and Caniglia 2020). Such collaboration is complex and can be challenging due to a variety of factors. Individuals across fields may not fully understand one another's vocabularies, what is considered data to individuals in one discipline may be seen as context information to individuals in another, and the approaches to solving a problem may appear divergent or even in contrast to one another across differing fields (Borgman et al. 2012; Freeth and Caniglia 2020; Mao et al. 2019). These sorts of problems can occur in any collaboration, and the number and complexity of complications only grows as interdisciplinary teams comprise more fields.

One relative constant across interdisciplinary collaboration teams is the presence of a statistician/data scientist, whether by official title or project role, as nearly every collaboration team desires to tell a story with quantitative data. A key tension is that statistical skills are necessary to tell these stories, yet the deepest understanding of the context of a project often lies with domain experts (aka subject matter experts or clients) who may not have such analytic skills (Zhang et al. 2020). The pervasive demand for statistics and data science analytical skills combined with the likelihood that the person on the team with these skills is often not the same person on the team with the deepest domain-specific knowledge strongly suggests the importance of training statistics and data science students to be adept interdisciplinary collaborators. This heterogeneity of skills and interests among the members of a collaborative team also argues for the importance of considering the domain experts' perspective when evaluating approaches to training statisticians and data scientists in interdisciplinary collaboration (Zhang et al. 2020).

Some prior work presents ways to teach interdisciplinary collaboration to statisticians and data scientists (Jeske et al. 2007; Thabane et al. 2008; Davidson et al. 2019; Sima et al. 2020; Vance 2021; Kolaczyk et al. 2021), though none of these describe the framework used for teaching collaboration skills. Other research focuses on evaluating the effectiveness of various approaches to teaching collaboration skills (Bangdiwala et al. 2002; Jersky 2002; Mackisack and Petocz 2002; Roseth et al. 2008; Metzger et al. 2023; Vance et al. in press) or on evaluating the effectiveness of a statistical consulting center (Estrada et al. 2021). These evaluation studies consider a variety of ways of teaching interdisciplinary collaboration but primarily rely on student self-reflections and a handful of Likert-scale items from domain expert (i.e., client) feedback surveys as evidence of pedagogical success. The lack of independent and impartial evidence for evaluating the effectiveness of teaching interdisciplinary collaboration, along with limited evidence from the perspective of the domain expert, is a gap in the existing literature.

The ASCCR Framework (Vance and Smith 2019) was designed to simplify the teaching and learning of collaboration by focusing on five principal components of interdisciplinary collaboration: **A**ttitude, **S**tructure, **C**ontent, **C**ommunication, and **R**elationship. We have found it to be effective and use it to educate and train our students in a mixed undergraduate and graduate capstone collaboration course. However, the effectiveness of ASCCR to train students for actual collaborations has only just begun to be evaluated.

In Alzen et al. (2023), we conducted a pilot study to begin evaluating the effectiveness of ASCCR for training novice statistics and data science students. We developed a rubric to score video recordings of collaboration meetings between statistical collaborators and domain experts. We compared a team of two novice student collaborators with an expert collaborator who met independently with one domain expert on one project. In this paper, we extend that study to 10 novice pairs on 10 collaborative projects to answer our overall research question:

> *RQ1: In what ways and to what extent is the ASCCR framework useful for training novice statistics and data science students to become effective interdisciplinary collaborators?*

In particular, we use video recordings of initial collaboration meetings, rubric scores of various components of the meetings, and domain expert feedback surveys[1] to answer our specific research question:

> *RQ2: During initial collaboration meetings with a domain expert, how well do novices perform compared to an expert collaborator?*

This paper proceeds as follows: In Section 2, we summarize the ASCCR framework we use to define effective collaboration. We describe the study context, people, design, and data in Section 3. In Section 4, we present mixed-methods results from 10 collaboration projects. We discuss the results and provide recommendations for instructors to incorporate ASCCR into collaboration, consulting, and capstone courses in Section 5. We conclude in Section 6.

## 2. ASCCR Framework for Effective Collaboration

In the current study, we used the ASCCR framework for collaboration skills (Vance and Smith 2019) to both train our students and evaluate the effectiveness of their collaborations. The framework comprises five components of effective collaborations: **A**ttitude, **S**tructure, **C**ontent,

[1] All study instruments, deidentified data, and code are available at https://osf.io/ajc7g.

**C**ommunication, and **R**elationship. The details of ASCCR are provided in Vance and Smith (2019) and elsewhere, but we provide a high-level summary to provide context for this study.

## 2.1 Attitude

The ASCCR framework considers three areas of a collaborator's attitude within a project: a generally collaborative work attitude, an attitude about themselves as contributors to the project, and their attitudes toward the domain experts as contributors to the project. The general attitude about collaboration centers on the idea that the goals of all members on the team are important and that shared goals developed through the collaboration are stronger than individually generated goals. The collaborators' attitudes toward themselves should be to recognize their abilities to provide a positive contribution to the overall project. Similarly, the idea of the collaborators' attitudes toward the domain experts is that the collaborators recognize the domain experts as equal intellectual partners on the project.

## 2.2 Structure

The structure of project meetings is central to the ASCCR framework because statistical collaborators can make the work easier for themselves and the domain experts if they structure the meetings in ways that allow all participants to focus their intellectual energy on the content of the meeting rather than on logistical and administrative tasks. This component of the framework adopts the POWER process for structuring meetings (Zahn 2019). Under this approach, the statistical collaborators show that they **P**repared for the meeting based on their awareness of the content area as well as connections they make between their prior work and interests and the current project. As the meeting **O**pens, the collaborator ensures that all team members share their expectations (i.e., what they want to accomplish) during their time together. Following, the collaborator facilitates the **W**ork time of the meeting toward addressing the agreed-upon goals. At the **E**nd of the meeting, the collaborator summarizes what was accomplished, plans for unmet expectations, and identifies next steps for all involved. More advanced collaborators also save time at the end of the meeting for the domain expert to **R**eflect on what went well during the meeting and what could be done differently next time to achieve greater success.

## 2.3 Content

For interdisciplinary collaborative projects to be most successful, statisticians and data scientists must ensure that they understand both the qualitative domain context and the quantitative statistical needs of the project (Leman et al. 2015; Ograjenšek and Gal 2016). Aspects of the project such as the context, data provenance, how the project problem fits within the larger field, and potential extensions of the work are all vital to understand prior to considering statistical analysis (Vance 2019; Zhang et al. 2020; Vance et al. 2022b). This information is important not only for appropriate understanding and modeling of the data, but also for translating statistical results back into the domain expert's field. This component of the ASCCR framework takes up the $Q_1Q_2Q_3$ approach to understanding and using data from collaborative projects (Vance et al. 2025). Upon joining a project, the statistical collaborator must first focus on the qualitative ($Q_1$) elements of the project before considering any quantitative ($Q_2$) analyses. After completing analyses, the statistics and data science collaborator then interprets the statistical results within the qualitative context of the project ($Q_3$) to co-create conclusions and recommendations with the domain expert and ensure that they can appropriately explain these results to others in their field (Olubusoye et al. 2021; Vance and Love 2021).

## 2.4 Communication

Communication across disciplines is one of the key complexities when considering interdisciplinary collaboration (Freeth and Caniglia 2020). Although some argue for specific project activities or even personnel to ensure appropriate "translation" across disciplines, the ASCCR framework suggests that statistics and data science collaborators can be trained to fill this role on a project through effective questioning, listening, paraphrasing, summarizing, and explaining (Vance et al. 2022a; d). The main goal in this communication is to ensure shared understanding of project goals, information, results, and implications for the field across project team members at each phase of the project (Vance et al. 2022a).

## 2.5 Relationship

The ASCCR framework closes with consideration of the relationship between the statistics/data science expert and the domain expert. Zhang et al. (2020) suggest that this relationship is vital for project members to gain shared understanding because the statistician/data scientist must be able to negotiate a relationship with the domain expert effectively in order to garner deep understanding of the project elements. Additionally, Vance (2020) suggests that a strong relationship should be an end goal of a collaboration. In addition, strong relationships are necessary for building sustained partnerships, and sustained partnerships are how some of the more valuable contributions to the field are possible. This is because long-term partnerships can result in statisticians with greater domain-specific knowledge as well as domain experts with increased statistical knowledge. As individuals develop deeper understanding of both fields, more opportunities for additional questions and investigations become possible (Blanchard 2018).

The ASCCR framework provides the theoretical lens through which we assess the effectiveness of statistical collaboration meetings in this paper in both observations of collaboration meetings as well as qualitative data collected from domain experts. This framework provides a thoughtfully developed standard for what determines effective collaboration and helps strengthen claims about the effectiveness of any process for teaching collaboration skills.

## 3. Study Context, People, Design, and Data

Our general research question is about evaluating the effectiveness of training novice collaborators with the ASCCR framework to become effective interdisciplinary collaborators. In subsection 3.1 we briefly describe our training program. Our specific research question is how novice students compare to an expert collaborator when conducting an initial meeting with a domain expert. These initial meetings were video recorded and evaluated by two raters using a rubric we previously developed. In Subsection 3.2, we describe the four sets of people involved in this study: novice student collaborators, the expert collaborator, the domain experts, and the two video raters. In Subsections 3.3 and 3.4, we describe our study design and the data.

### 3.1. Context of our Training Program

The Laboratory for Interdisciplinary Statistical Analysis (LISA) is a statistical collaboration laboratory created in 2016 at the University of Colorado Boulder, a research intensive university. LISA's externally facing goal is to increase the quantity and quality of statistics and data science applied to high-impact research at the university and in the community. Internally, LISA's goal is to train the next generation of collaborative statisticians and data scientists to become effective interdisciplinary collaborators. Undergraduate and graduate students (in approximately equal numbers) join LISA by enrolling in the course titled *Statistical Collaboration*, during which they learn and practice the collaboration skills in the ASCCR framework and collaborate on three LISA projects. Projects primarily come to LISA via a public website through which indivdiuals from the university and surrounding community provide basic information about their project and their desired statistical collaboration help.

Undergraduates in *Statistical Collaboration* are typically final-year statistics and data science (SDS) majors who have taken courses such as probability, R, python, mathematical statistics, regression, statistical learning, and statistics electives such as time series, spatial statistics, or Bayesian statistics. Graduate students in *Statistical Collaboration* come from a variety of STEM and social science M.S. and Ph.D. programs. As prerequisites, they would have taken at least two graduate level SDS courses covering probability, inference, regression, and statistical learning techniques in R. Many would have taken additional statistics courses in their programs of study and/or as undergraduates.

Students in *Statistical Collaboration* are deemed ready for their first collaboration project at around the fourth week of the semester after acquiring a basic understanding of the ASCCR framework. They achieve this through a systematic process of preparation and practice, and then are supported during and after their collaborations through self-reflection and mentoring (LeBlanc et al. 2022). Specifically, before their very first collaboration meetings, students read about the ASCCR frame (Vance and Smith 2019), the POWER structure (Zahn 2022), and the $Q_1Q_2Q_3$ workflow (Vance et al. 2025); observe and reflect on a real LISA collaboration meeting; and practice some of the ASCCR skills in class using role plays (Davidson et al. in press). For the remainder of the course, students continue learning and practicing ASCCR skills and discuss their collaboration projects in class. A more detailed description of this course and our training process is available in Vance and Pruitt (2022) and Alzen et al. (2024).

Depending on student demand, a course titled *Advanced Collaboration* is offered to students who have previously taken *Statistical Collaboration*. In *Advanced Collaboration*, students continue advancing on their journey from novice to expert (Alzen et al. 2024) by refining their ASCCR skills and collaborating on three or four additional LISA projects. In addition, every student is a focus of at least one Video Coaching and Feedback Session (VCFS), in which the entire class reviews three short clips of a video-recorded collaboration meeting and provides feedback and coaching to the students in the video. During the period of this study (2021-22), LISA conducted 12 VCFS. *Statistical Collaboration* was taught in Fall 2021 and Spring 2022, and *Advanced Collaboration* was taught in Spring 2022. All three of these courses were taught by this paper's last author.

## 3.2. People of the Study

### *3.2.1. Student Collaborators (Novices)*

Twenty students (12 undergrad and 8 graduate) took *Statistical Collaboration* in Fall 2021, and 12 (5 undergrad and 7 graduate) took it in Spring 2022. *Advanced Collaboration* was taught in Spring 2022 to 14 students (8 undergrad and 6 graduate). All undergraduates in the courses were SDS majors; several were double majors, with additional majors including computer science, psychology, and political science. The graduate students were widely dispersed among at least 10 departments, with students coming from psychology, integrated physiology, mathematics, materials science and engineering, and the other departments listed below.

Sixteen unique students (9 undergrad and 7 graduate) filled the 20 novice collaborator slots on the 10 focal projects in our study. Half of these projects were led by undergraduates and the other half by graduate students; one graduate student led two projects. Approximately 40% of the students in the collaboration courses and in our study sample were female. The novices had collaborated on between 0 and 5 projects prior to the focal project (mean = 2.1, median = 1.5). Four novices had their very first collaboration project included in this study (Projects 5 and 9). The initial meetings for those two projects occurred during Week 4 of the semester. Table 1 lists the novices' departments, levels, and their number of prior projects.

**Table 1.** Novice collaborators' departments, levels, and numbers of prior projects.

| ID | Department | Level | Lead | Project # | # Prior Projects |
|---|---|---|---|---|---|
| 1101 | SDS/Computer Science | Undergrad | X | 7 | 2 |
| 1107 | SDS/Computer Science | Undergrad | | 7 | 2 |
| 2119 | SDS | Undergrad | X | 2 | 3 |
| 2111 | SDS | Undergrad | | 2 | 4 |
| 1204 | SDS | Undergrad | | 8 | 1 |
| 2108 | SDS | Undergrad | X | 6 | 4 |
| 1205 | SDS | Undergrad | | 9 | 0 |
| 2106 | SDS/Psychology | Undergrad | X | 4 | 5 |
| 2120 | SDS/Computer Science | Undergrad | X | 10 | 1 |
| 1202* | Applied Math | Graduate | X | 1, 5 | 1, 0 |
| 1201 | Applied Math | Graduate | | 1 | 1 |
| 1206 | Biochemistry | Graduate | X | 9 | 0 |
| 1211* | Biomedical Engineering | Graduate | XX | 5, 8 | 0, 1 |
| 2105* | Aerospace Engineering | Graduate | | 6, 10 | 4, 1 |
| 2114* | Astro. and Planetary Sciences | Graduate | X | 3, 4 | 4, 5 |
| 2113 | Speech, Lang., and Hearing Sciences | Graduate | | 3 | 4 |

* Student worked on two projects in the study.

### *3.2.2. Expert Collaborator*

The expert collaborator in our study was a recent biostatistics Ph.D. graduate. At the beginning of our study, she had three years of consulting and collaboration experience. She had worked on many short-term consultations and approximately 15 collaborative projects at statistics labs at two research universities. By the end of the study period, she had a fourth year of experience and had worked on approximately 30 collaborative projects in total.

### *3.2.3. Domain Experts*

There were 37 LISA projects conducted by pairs of novice students in Spring 2022. In this population of projects, 15 (41%) came from graduate students, 9 (24%) from faculty, 8 (22%) from undergraduates, 3 (8%) from community members, and 2 (5%) from campus staff. Eight of the 10 focal projects in our study came from graduate students and the remaining two from faculty (see Table 2). Descriptions of the content of these projects are available in our data repository at https://osf.io/ajc7g.

**Table 2.** Domain expert roles and affiliations for the 10 focal projects of this study.

| Project # | Role | Department | Project Content |
|---|---|---|---|
| 1 | Faculty | Classics | Sex ratios among ancient slaves |
| 2 | Graduate Student | Integrative Physiology | Stress modulation of circulating miRNA |
| 3 | Faculty | Anthropology | Characterizing variation in lemur behavior |
| 4 | Graduate Student | Psych/Neuroscience | Effects of cannabidiol treatment on anxiety |
| 5 | Graduate Student | Computer Science | Estimating blocked resources |
| 6 | Graduate Student | Mechanical Engineering | Improving cartilage implants |
| 7 | Graduate Student | Environmental Studies | Mitigating human-wildlife conflict |
| 8 | Graduate Student | Geography | Channel morphology and flood inundation |
| 9 | Graduate Student | Linguistics | Verb-specific biases in Tagalog |
| 10* | Graduate Student | Geological Sciences | Terrestrial paleoclimate |

* Focal project for the case study in Alzen et al. (2023)

### *3.2.4. Video Raters*

Two video raters (the first and third authors of this paper) rated the initial meeting video recordings of the 10 focal projects in our study according to a rubric (see Appendix A) developed for this project in the previous year. During development and validation of the rubric, the two video raters—along with domain experts from five different disciplines—observed and scored four collaboration videos to develop the rubric for use across fields (Alzen et al. 2023). Individuals independently scored videos, came together to compare scores, and improved rubric language to clarify and define meaning across the scoring range (1-minimal to 4-mastery) and applicability, regardless of project domain.

For this study, the two video raters first watched the full videos individually, took notes on what they observed, and used a checklist for elements of the ASCCR framework (e.g., collaborators ask the domain expert about their "wants" for the meeting). The raters then used the checklist, notes, and extra viewings of the videos when necessary to identify rubric scores for 15 of the 18 elements on the rubric; they did not score the three elements that are typically applicable only later in a collaboration project (i.e., Content: $Q_2$, Content: $Q_3$, and Communication: Explanation). Following individual scoring, the raters met to reach consensus and validate codes (Miles et al. 2020). They averaged items for the components with multiple items to yield five overall ASCCR component scores for each video.

## 3.3. Study Design

To help evaluate the effectiveness of our education and training program, we designed a study to directly compare the collaboration effectiveness of novice student collaborators vs. an expert collaborator. The process of "doubling" initial collaboration meetings, video recording both of them, completing the project, soliciting feedback from the domain expert, and using our rubric to evaluate the videos involved 11 steps (see Figure 1). The process started when a domain expert submitted a collaboration request to LISA (step 1). The instructor for the collaboration courses then shared the project details with the students in class (step 2). Based on students' interest in the project, which students needed additional projects to fulfill the course requirements, and the instructor's recommendations, this paper's third author matched two students to collaborate on each project (step 3) and then offered the option to the domain expert to participate in the study (step 4). All individuals involved in the project meetings completed informed consent forms approved by the University of Colorado Boulder's Institutional Review Board (Protocol 18-0554). The instructor was blinded to the students' and domain experts' consent to participate in the study. The two novice students and the expert collaborator coordinated to schedule two initial meetings with the domain experts who had consented to participate in the study (step 5). These meetings were scheduled on different days within one week of each other, typically two days apart. The domain experts' lack of consent to participate or inability to schedule two separate initial meetings were the main reasons a project was not included in our study. In those cases, the domain expert met only with the two student collaborators and their initial meeting was not video recorded.

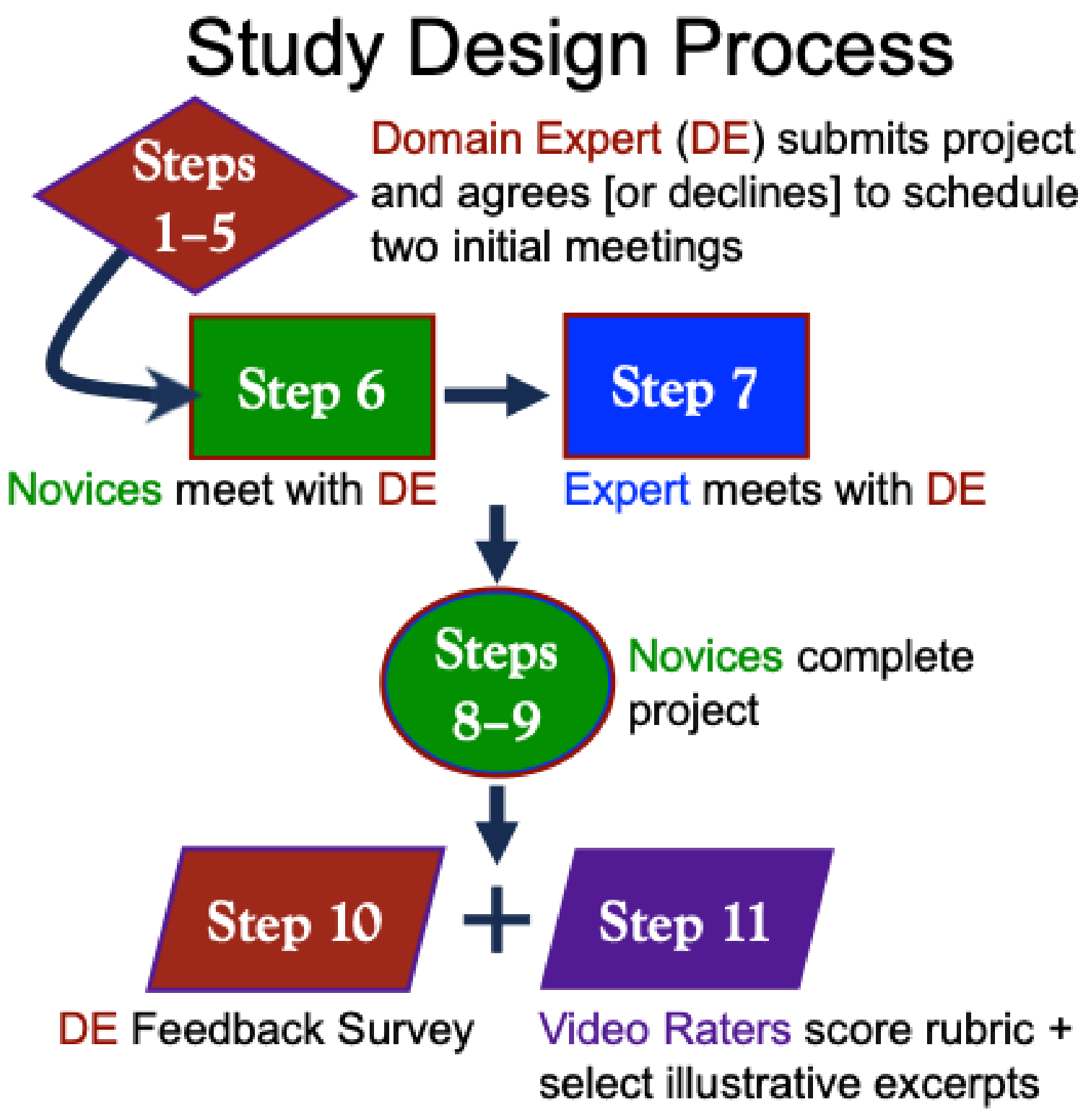

**Figure 1.** Flowchart of the 11-step study design and data collection process. Domain experts meet with two novice collaborators and an expert collaborator. Meetings are video recorded. This process produces domain expert feedback survey data, video rubric scores, and illustrative excerpts from the videos.

For the projects remaining in the study, the domain experts first met with the team of two novice collaborators on Zoom (step 6), and this meeting was video recorded. Following, the domain experts had a second "initial" Zoom meeting with the expert collaborator (step 7), also video recorded. The primary purpose of these initial meetings was for the domain expert to explain the project, give a general orientation to the statistics or data science need, and for the collaborators to ensure they understood the project before conducting or providing advice on statistical analyses. By design, the domain experts always met with the expert collaborator second. This choice likely biased the outcomes of the initial meeting *in favor* of the expert collaborator over the novices because after the first initial meeting with the students, the domain experts had practice describing their projects and needs. They had a better idea of how to succinctly describe the overall research design, data, and relevant questions as well as a sense of what questions might arise for a newcomer. Since the students had a disadvantage in the initial meeting, this strengthens potential evidence of the novices' ability to conduct collaboration meetings well compared to an expert.

Following the second initial meeting, the expert collaborator conferred (typically via email) with the novices to share notes, answer questions, or provide guidance on the project (step 8). The students continued meeting with the domain expert—typically twice more—until the project was completed (step 9). Then the first author solicited feedback from the domain expert via a survey (step 10). Finally, the two video raters used the rubric to evaluate both of the initial meeting video recordings and selected illustrative excerpts from the videos (step 11).

We started this process mid-semester in Fall 2021, and got only one project that had two usable initial meeting videos and domain expert feedback. That project was used for the case study in Alzen et al. (2023). This design yielded nine more projects in Spring 2022.

### 3.4. Data

Our study design yielded three types of data from 10 collaboration projects: rubric scores from eight[2] pairs of initial meeting videos, illustrative excerpts from AI-generated transcripts of these videos, and feedback from 10 domain experts on end-of-project surveys. As described above, the video raters used the rubric (see Appendix A) to create overall component scores for Attitude, Structure, Content, Communication, and Relationship for each video. They also selected passages from the transcripts to illustrate the results presented in Section 4.

At the conclusion of the project (typically the end of the semester), we collected feedback from the domain experts regarding their collaboration experiences with the novice collaborators and the expert collaborator[3]. The survey (see Appendix B) included three items about the effectiveness of the collaboration process and five items about the outcomes of the collaboration. We also asked four open-ended questions that touched on both aspects.

Specifically about the collaboration process, the domain experts were asked how much they agreed on a six-point agreement scale from “Strongly disagree (1)” to “Strongly agree (6)” that they “felt welcomed during meetings” and had “positive interactions outside of meetings.”

[2] The video raters were unable to evaluate two of the projects due to technical problems with the video recordings.
[3] One of the domain experts provided feedback for the novices only. Another domain expert did not complete the last two Likert-scale items for the expert collaborator.

They were also asked to rate the "usefulness of meetings" on a six-point scale from "Not at all (1)" to "Extremely (6)".

To evaluate the outcomes, we asked domain experts (on the six-point agreement scale) if they found the collaboration to be "helpful for their project" and if they were "satisfied" with the experience. We also asked them to rate on a six-point scale from "Very Low (1)" to "Very High (6)" the collaborators' "overall contribution to the project" and the "strength of their relationship with the collaborators." Lastly, we asked a net promoter score question (Reichheld 2003) for domain experts to indicate on an 11-point scale from "Not at all Likely (0)" to "Extremely Likely (10)" how likely they were to recommend LISA to a friend or colleague.

## 4. Results

To answer our specific research question about how well novices perform during initial collaboration meetings compared to an expert, we compared the rubric scores on eight pairs of video recorded initial collaboration meetings. Table 3 provides a high-level overview of the average ASCCR component scores. The mean scores of the expert and novices were similar (within 1 SD) for Attitude, Structure, and Relationship. However, the expert appeared marginally stronger on the Content and Communication components of the ASCCR framework. Table 3 also shows that the mean scores are not substantially different based on the novice pairings. Undergrad-Undergrad (U2), Grad-Undergrad (GU), and Grad-Grad pairs (G2) had similar mean rubric scores.

**Table 3.** Mean (SD) rubric scores by component (4-point scale) by type of collaborator and novice pair.

| Component | Expert | Novices | U2 | GU | G2 |
|---|---|---|---|---|---|
| Attitude | 4 (0) | 3.75 (0.34) | 4 (0) | 3.84 (0.19) | 3.34 (0.47) |
| Structure | 3.38 (0.32) | 2.87 (0.53) | 2.33 (0) | 3.12 (0.51) | 2.92 (0.59) |
| Content | 3.12 (0.83) | 1.88 (0.99) | 1.5 (0.71) | 2.25 (1.26) | 1.5 (0.71) |
| Communication | 3.91 (0.13) | 2.78 (0.84) | 2.12 (0.53) | 3.06 (0.94) | 2.88 (0.88) |
| Relationship | 3.62 (0.52) | 3.12 (1.13) | 2.5 (2.12) | 3.25 (0.96) | 3.5 (0.71) |
| N | 8 | 8 | 2 | 4 | 2 |

In the following subsections, we show the rubric scores for the pairs of meetings and draw on video transcripts to characterize the differences and similarities between the expert and novice collaborators. We start with Communication because we saw the largest differences between novices and the expert for that component of ASCCR. In Section 4.4, we augment these results with an analysis of domain expert feedback on the projects.

## 4.1. Communication

The main purpose behind the Communication component of the ASCCR framework is for the collaborator to ensure that they and the domain expert have a shared understanding about the project. This is evidenced by the ways the collaborator listens, asks questions, paraphrases, and summarizes during meetings. Although all novices used each of these skills in their meetings, we generally saw the expert do so with more regularity and depth, as shown by the Communication rubric scores in Figure 2. The projects (y-axis) are ordered from the most experienced novice pair (Project 4) to the least experienced pair (Project 9). We saw no association between the average Communication rubric score (x-axis) and the novice pair type or their prior experience.

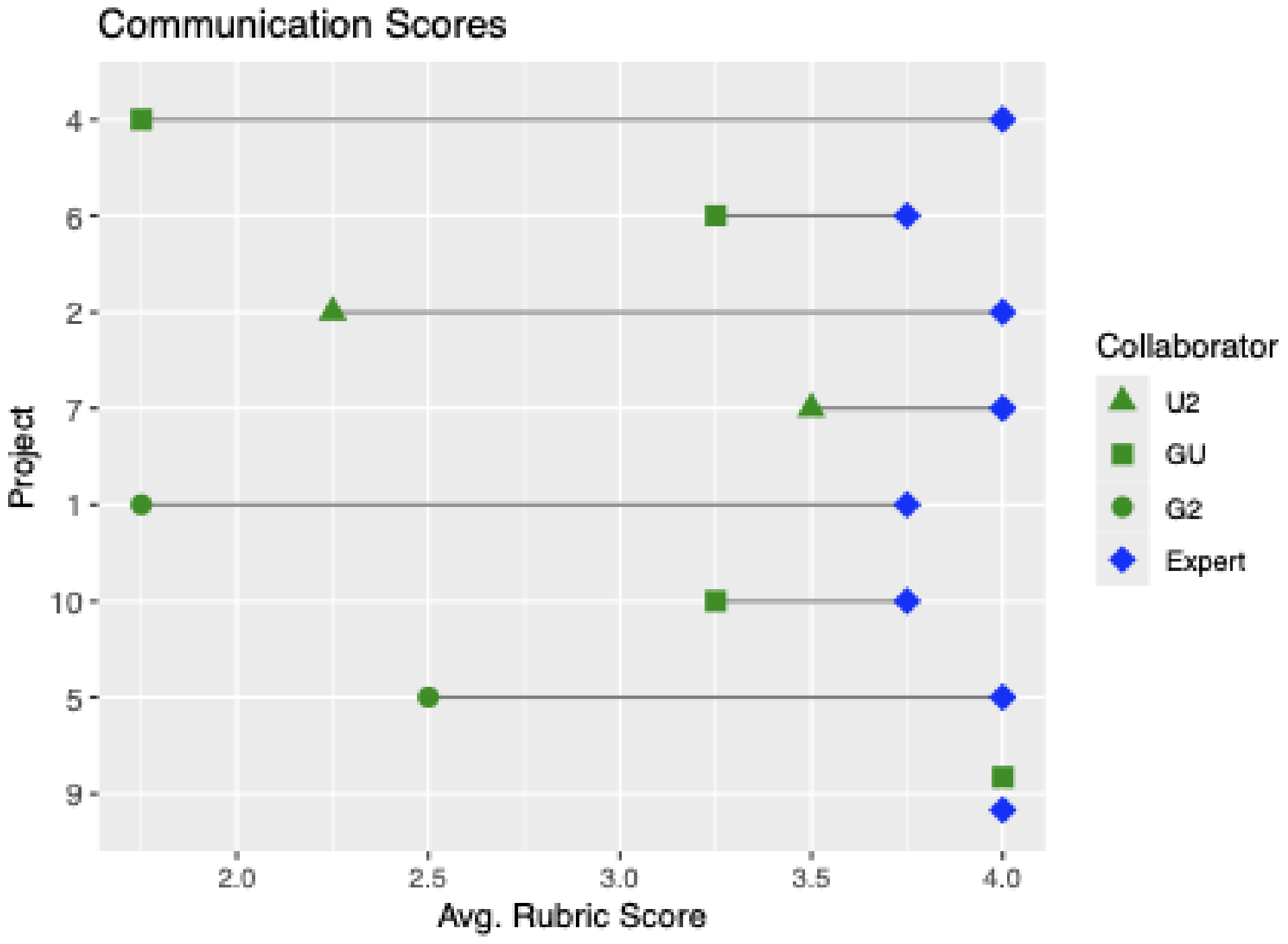

**Figure 2.** Average Communication rubric scores by collaborator type. The plot compares expert performance (blue diamonds) against three classifications of novice collaborators: U2, GU, and G2 (green shapes). Projects are ordered from the most experienced novices (top) to the least experienced (bottom).

As an example, in Project 1, after the domain expert gave an introductory summary of his research, he said, “I’ll pause and see if you have anything you wanted to follow up on or questions.” The novice students indicated they were taking notes while he spoke, asked if he could see their notes, and then asked the domain expert to continue talking. The domain expert asked a second time, “I was waiting if you had any questions.” In response, the novices asked a question about the data sources for the project, and the domain expert proceeded to talk for about 20% of the meeting about data sources, during which the students did not pose any other questions. At that point, the domain expert asked if the students would like to see a picture of the historical artifact that serves as a data source he had been discussing. The students responded with “Yeah. Sure.” Even with the domain expert providing multiple opportunities for the collaborators to ask questions, they did little to ensure they fully understood the details of the project or communicate their learning about the project back to the domain expert.

Toward the end of the meeting, the students began using more of their communication skills. They asked things like “Could you elaborate a little more on…”, “I had another question about…”, and “Is that what you’re looking for?” However, these communication efforts were rather dissimilar from the expert collaborator on the same project. When the expert collaborator met with the domain expert for Project 1, the turns of talk were much different. Nearly every time the domain expert shared information about his project, the expert collaborator checked her understanding by summarizing, paraphrasing, or asking a follow-up question such as “Is that what you were saying?” or “Am I understanding that part of your question is…” or “What do you mean?” Additionally, the expert collaborator actively asked questions about projects rather than solely waiting for the domain expert to share (e.g., “Why is [that] important to the study?” and “When you say […] what do you mean?”). As a result, the expert received higher rubric scores on Communication because she appeared to understand much more about the context of the project and the details of the research questions the domain expert was trying to answer.

## 4.2. Content

Content was scored with a single rubric item since we only observed the initial collaboration meeting for each project. As seen in Figure 3, the expert generally scored higher than the novices. We saw no association between the average Content rubric score and the novice pair type or their prior experience.

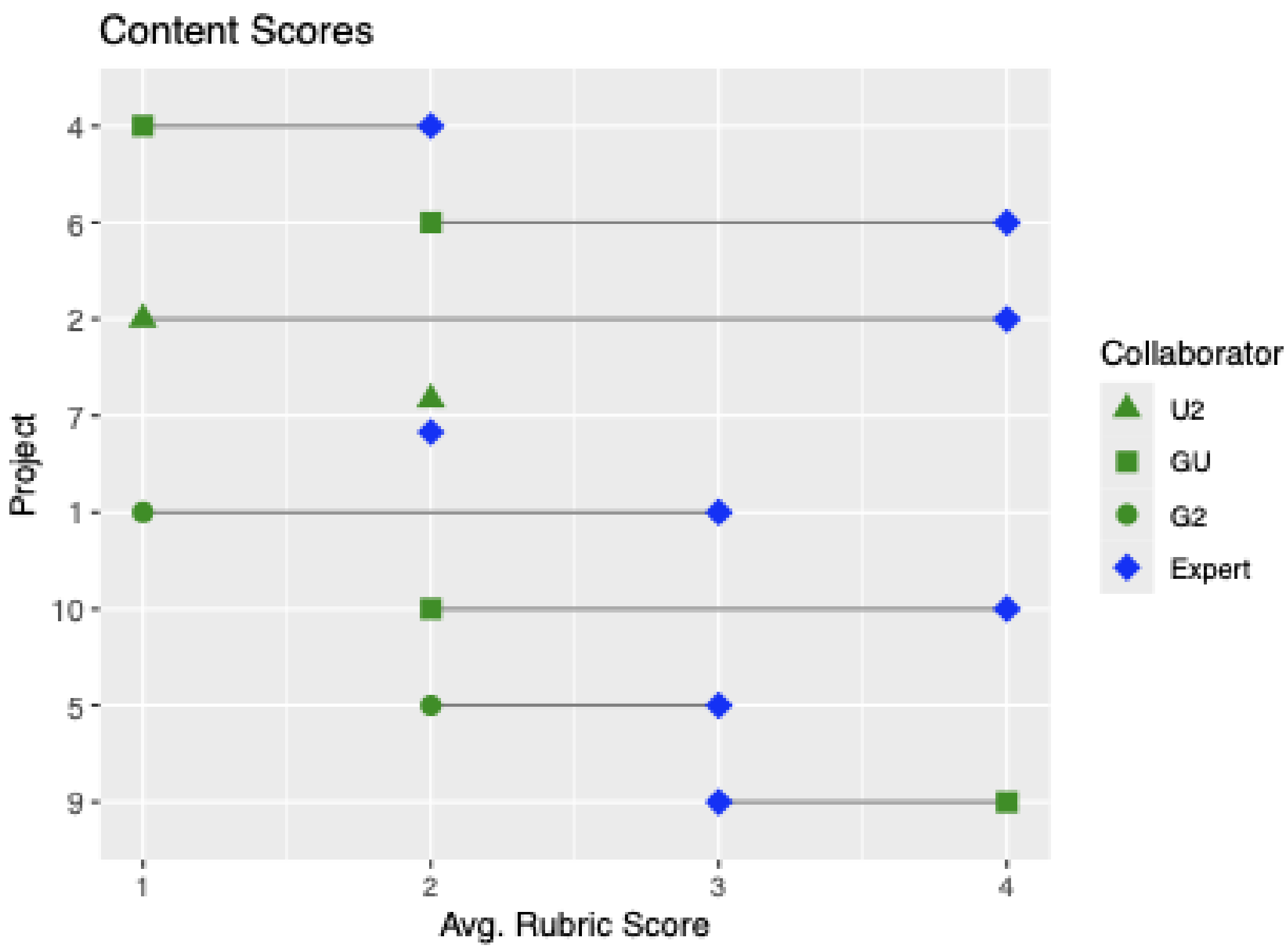

**Figure 3.** Content ($Q_1$) rubric scores by collaborator type. The plot compares expert performance (blue diamonds) against U2, GU, and G2 novice collaborators (green shapes). Projects are ordered from the most experienced novices (top) to the least experienced (bottom).

Qualitative differences in the Content component of ASCCR between the expert and novice collaborators were more subtle than in the Communication component. Both the novices and the expert began initial meetings soliciting contextual information about the projects, but the degree of detail varied between the two. For example, in Project 2, the novices began by asking the following:

> What we want to get done from our end is kind of to learn more about your goals—why you care about this project and what sort of statistical methods you envisioned. And then obviously we'll talk to you about what we think. So that's from our end, what we want to get done. Do you have any specific wants other than that?

Conversely, the expert began her meeting for Project 2 as follows:

> I'd love to just have you give me an overview of your research. So what's the big picture question and sort of the real world implications of your research. What your hypothesis and question is. If you know what your outcome and predictor is, which based on your submission to LISA it looks like you do, so we can talk about that, which is great. What does your data or will it look like if you have data or if you're going to collect it, what will it look like? And then why you've come to LISA. And then also talking about what deadlines you have for this project. So those are the four big things I have. Is there anything you want to add to this meeting agenda?

In each of these cases, the collaborators framed the discussion around the domain expert's goals for the project. The expert collaborator emphasized a desire for a more detailed understanding of the project and did not mention sharing her own ideas about what should happen in the project. Her initial focus was on understanding the context and left any statistical contributions she might have for the project until later.

The contrast in the approaches between the expert and the novices persisted through this project's initial meetings. During the novice meeting, the domain expert screenshared data approximately three minutes into the meeting after sharing some very high-level context of the study. He moved away from the data to continue talking about the project, but the students asked him to return: "Would you mind going back to the data really quickly? I just want to make sure I was understanding all of that." Although a novice student mentioned wanting to understand the data and did not immediately start talking about how to use the data, he did not yet have a full understanding of the general concepts in the project before he jumped in to trying to understand the data.

Conversely, the expert collaborator purposefully held off on discussing data or statistical analysis until she gained a strong understanding of the project context. We saw this multiple times:

> "I also want to make sure I understand your data fully before we dive too much into the different methods. […] You had how many [subjects] in control and how many not control? […] And how do you measure your outcomes?"

A little later in the meeting, the domain expert offered his code to the expert collaborator, but rather than looking she said: "Before we [look at your code]. Do you know if any of [your variables] are highly correlated with one another?" This is a clear example of how the expert collaborator put more effort into understanding the full qualitative context of a study before beginning to think about the quantitative statistical work.

## 4.3. Attitude, Structure, and Relationship

The rubric scores in Figure 4 indicate that—unlike for Communication and Content—the expert and novice collaborators performed relatively similarly for Attitude, Structure, and Relationship, with no associations between the average component scores and the novice pair type or their prior experience.

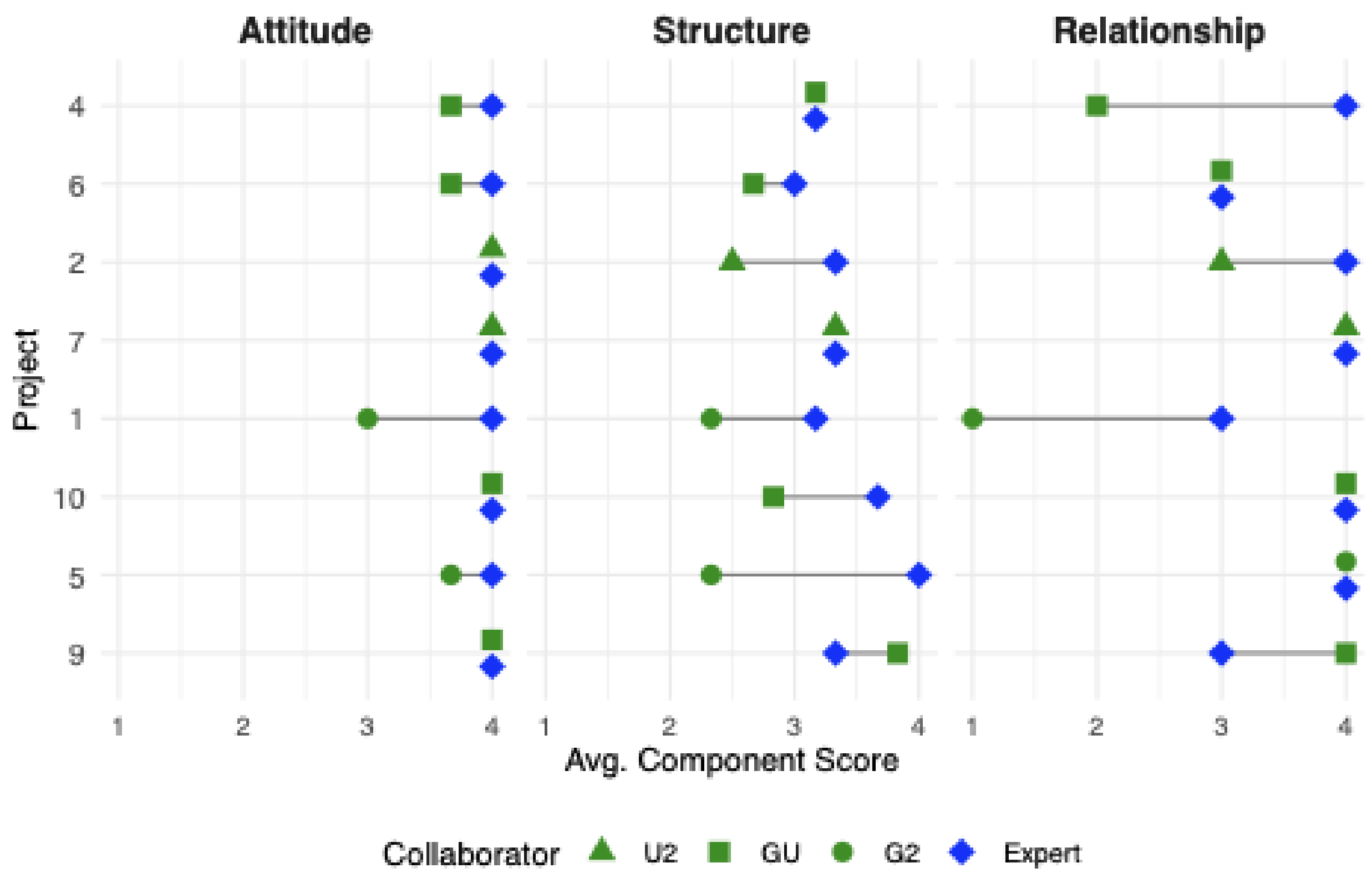

**Figure 4.** Average rubric scores by collaborator type across three components: Attitude, Structure, and Relationship. Scores are plotted for individual projects and compare expert performance (blue diamonds) against U2, GU, and G2 novice collaborator types (green shapes).

Attitude is difficult to observe as this is more of an internal latent construct. However, as the Attitude component of ASCCR emphasizes the importance of collaborators having positive attitudes about themselves (including an attitude of intending to be helpful), the domain experts, and the work the teams can do together, we attempted to identify these elements of Attitude through the ways collaborators acknowledged all individuals in the meeting as experts with valuable contributions to the project and attitudes of helpfulness. For example, in Project 7, the novices made several comments to the domain expert about the work she had already done on the project. Prior to coming to LISA, the domain expert found some code online and attempted to apply it to her project. When going over the code, the students said things like "you're doing great" and "you're killing it." Similarly, the expert collaborator replied "you're doing some crazy [advanced statistical analysis], which is really cool. That's awesome" and "[your analysis] seems really interesting and useful, and I like the visualizations you made with it." These excerpts show how all collaborators recognized the domain expert as a valuable contributor to the technical aspects of the project and not solely reliant on LISA support for statistical analysis.

In addition to comments about the work that domain experts completed in statistics, the collaborators also recognized the helpful contributions LISA could make to a project. After the domain expert from Project 2 explained his work and some of the analysis he had already completed, the novices said, "It really looks like you have an awesome start there." Similarly, the expert collaborator responded by saying, "You're crushing it." Following these comments, all collaborators additionally specified the value LISA could bring to the project. For example, the novices said:

> We'll make sure everything's been done appropriate[ly], that we're not dealing with a violation of some other assumption along the way you might have missed […] It would be lovely if you would send us what you have as a start though, because it looks like you really do have a solid start.

The expert collaborator responded to a parallel moment in Project 2 quite similarly: "I'm glad that you came to us so that we can help fill in the remining gaps [in your analysis]."

Something notable is that the expert collaborator received the highest score on the rubric (4) for Attitude for every video in the sample. Although none of the domain experts explicitly named her attitude in their feedback survey responses, it was obvious they did appreciate her attitude based on their comments. For example: "Meeting with [expert] was pleasant and a positive experience" (Project 10) and "[Expert] took the time to truly understand my project and the data to make recommendations on the stats and models I was using" (Project 8). The domain expert for Project 3 (DE3) went as far as to compare the expert and novice collaborators: "[Expert] was easier to build initial rapport with [than novices]." Although DE3 did not explain specifically why the expert collaborator was easier to build rapport with, this an indication for why the expert received slightly higher scores on Attitude than the novices.

Consistent implementation of the Structure component of ASCCR was obvious across meetings as well. All collaborators regularly used the POWER structure for organizing their collaboration meetings with the domain experts. The explicit POWER format for organizing meetings appeared to be an easy tool for the collaborators to adopt into their meeting facilitation. This was most obvious in starting meetings with sharing notes, having conversations about the project and meeting goals, and ending the meeting by summarizing what occurred and making a plan for next steps.

In Project 4, all collaborators sent meeting notes to the domain expert ahead of time and mentioned them at the top of the meeting. The novices said, "I just put a meeting note [link] in the chat again so everyone can look at it," and the expert collaborator asked, "Did you get the meeting notes that I sent just a little bit ago?" Similarly, in both meetings, the collaborators asked some form of, "We have the meeting scheduled until [time]. Does that still work for everybody?" Finally, at the end of the meeting, collaborators ensured the work was finished on time and that everyone in the meeting knew the plan for next steps. For example, the novices summarized what occurred during the meeting and revisited the goals that had been discussed at the beginning of their time together. "Just to recap a little bit, [summary of discussion]. Okay. I think we've been through everything we set out to do, so I suppose we can wrap it up here."

Finally, the novice and expert collaborators also appeared to handle the Relationship component of ASCCR comparably across projects. Similar to the Attitude component, it is difficult to observe elements of a positive relationship. However, we looked for evidence of the collaborators and the domain experts commenting positively about the meeting. This type of evidence occurred in several meetings. For example, DE7 made comments such as "I'm excited to have so much support from you guys [novices]" and "I'm just grateful for your [the expert's] help because I did not feel confident about this at all, and now I feel so much better." DE4 said, "It's been super helpful and I'm excited to tackle this now, so I appreciate y'all [novices]." Similarly, collaborators made comments such as: "It's a really interesting project. Thank you for explaining [it] so thoroughly" (Project 1 expert collaborator), "Thank you so much for giving us this background, for showing the data. It was very helpful. I think we have a better grasp on this now" (Project 6 novices), and "We're really excited [to be] on this project. It's really interesting" (Project 7 novices).

Another way collaborators work to build relationships with the domain experts is to make connections between their background and that of the project. This happened with comments such as: "I come from a biomedical engineering background, so I have some knowledge of biological data and stuff, and I'm very interested in this project" (Project 6 novice) and "I did my undergrad in applied math and I'm very excited about your project because I saw you're in mechanical engineering and you're talking about 3D printing and stuff, so that's really cool" (Project 6 expert). Collaborators also worked to strengthen the relationship when making positive comments about a project even if they had little to no prior experience in the field. For example, "I got my degree in statistics, and I have very little classics background, but I'm very interested in learning, so I'm very excited about your project" (Project 1 expert).

## 4.4. Evidence from Domain Expert Feedback Surveys

### *4.4.1. Process of Collaboration*

Domain experts rated the novice and expert collaborators similarly on the process of collaboration. On six-point scales, domain experts rated how welcomed they felt during meetings, the usefulness of the meetings, and their interactions outside of meetings. Figure 5 shows diverging bar charts of the predominantly positive responses from the domain experts on these *process* items, with the novices receiving slightly higher ratings than the expert.

**Figure 5.** Diverging bar chart showing the domain expert feedback on three process items: “Felt welcomed”, “Useful meetings”, and “Positive interactions outside of meetings.”

### *4.4.2. Outcomes of Collaboration*

All domain experts “Somewhat agreed (4)”, “Agreed (5)”, or “Strongly agreed (6)” that both sets of collaborators were *helpful* and that they were *satisfied* with their LISA experiences.

On average, the novices scored slightly higher than the expert. The mean (SD, n) ratings for the novices on the *helpful* item was 5.30 (0.67, n = 10) compared to 5.00 (0.87, n = 9) for the expert. On the *satisfied* item, the mean novices rating was 5.50 (0.53, n = 10) compared to 5.33 (0.71, n = 9) for the expert. The domain experts also rated the collaborators' *contribution* to the project and the strength of their *relationship* with the collaborators. The mean novices rating for *contribution* was 3.70 (0.95, n = 10); the expert's mean rating was 3.50 (1.20, n = 8). These mean scores fell in between the "Low (3)" and "High (4)" rating on a six-point scale. Somewhat higher, the mean novices rating for *relationship* was 4.50 (0.71, n = 10), and the expert's mean rating was 3.88 (0.99, n = 8) on the same six-point scale. Figure 6 shows the distribution of these domain expert feedback scores on the outcomes of the collaboration.

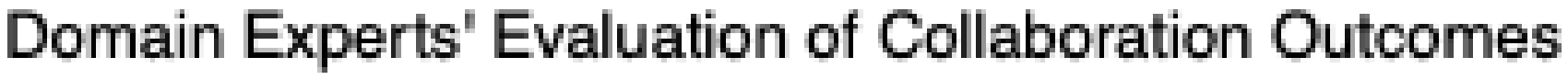

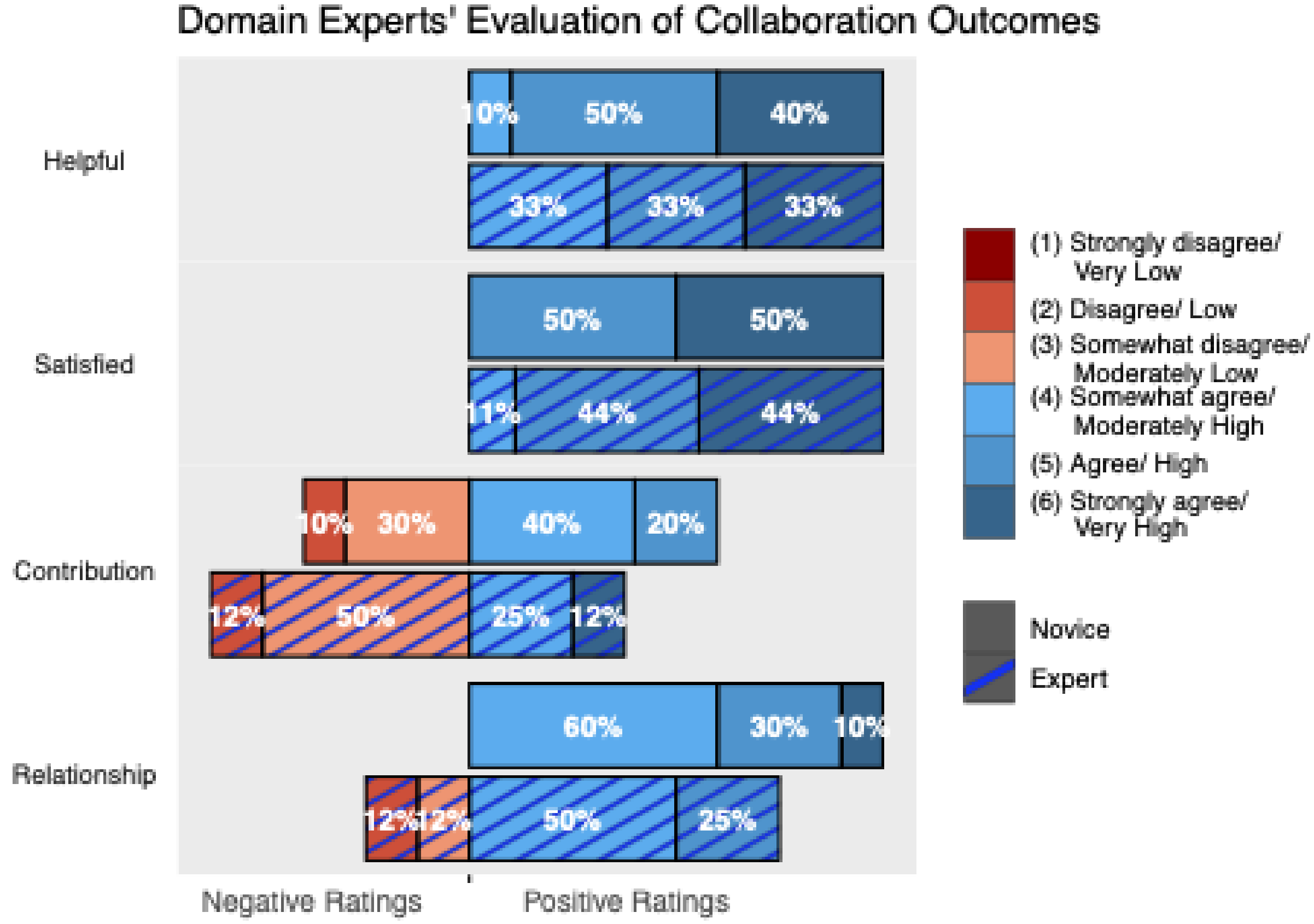

**Figure 6.** Diverging stacked bar chart of domain expert evaluations across four collaboration outcome dimensions: helpfulness, satisfaction, contribution, and relationship. Responses are recorded on 6-point rating scales ranging from Strongly disagree/Very low to Strongly agree/Very high.

Finally, we asked domain experts to rate how likely they were to recommend LISA to their friends or colleagues on a scale from 0 to 10, qualitatively ranging from "Not at all likely (0)" to "Extremely likely (10)". Both sets of collaborators engendered similarly high scores from the domain experts. The mean novice score was 8.80 (1.32, n = 10); the mean expert score was 9.00 (0.87, n = 9).

*4.4.3. Open-ended Comments*

We lastly draw on themes from the domain expert feedback survey comments to understand more about the domain expert experiences working with LISA. The open-ended survey items asked domain experts to give specific feedback on collaborators, including naming what they did well (i.e., strengths) and ways they could improve. Domain experts were also asked how working with LISA influenced their decision making and research as well as any additional comments or general feedback they had about LISA. An overarching and persistent theme across the open-ended items was general satisfaction with the collaboration process. In addition, domain experts specifically named communication and boosted confidence in their own statistical work on projects as valued outcomes from working with LISA collaborators.

Domain experts generally indicated appreciation for the effort collaborators put in to understanding their research and data. For example, DE10 shared that the novices "were very respectful and clearly were taking time to understand aspects of my research most relevant to the statistical goals." Similarly, DE9 reported the following:

> [The novices and expert collaborators] were able to adapt and learn quite a bit about my topic in a short amount of time. It's not easy!! They were good at setting a plan, asking questions, and helping me get to the heart of the questions that I was interested in. […] Overall, I think they were great to collaborate with.

More specific comments around communication included "the [novice] collaborators were great at communication both during the meetings we had and over email" (DE8) and "[the novice] understood my questions/research with seeming ease (and in spite of my inarticulateness); he also wrote very helpful emails explaining some of the suggested methods and follow up" (DE3). Half of the domain experts made specific unsolicited comments about how their own statistical confidence increased because of collaborating with LISA. DE6 indicated he "was more confident in analyzing [his] data;" DE8 stated, "My work with LISA helped me better understand statistical models;" and DE2 reported, "LISA helped give me statistical confidence."

Finally, DE3 made comments that revealed how our research design influenced her experience with collaborators. She reported:

> [The novices] worked thoughtfully to understand [my] research and data (and challenges). Both provided unique perspectives on approaches to data and also approached meeting with me with a lot more empathy than I expected based on previous interactions seeking statistical help from both within and outside of LISA.

However, DE3 also stated the following:

> I met with [expert] after [the novices], and I think my introductory meeting with her was ultimately more successful than my initial meeting with [the novices] for two reasons 1) I got practice with my meeting with [the novices] about what things to emphasize about my research so I felt that I was more equipped to talk about things with [expert] and 2) [expert] was able to ask more specific targeted questions about my data and my research.

We anticipated experiences like that reported by DE3. In some ways, this was a purposeful design element of the research project in order to give the novices the most authentic collaboration meeting experiences and to ensure that the research design biased outcomes in favor of the expert collaborator to help strengthen claims about the effects of using ASCCR to teach collaboration. However, a larger study that could randomize the condition of which collaborator held the first meeting with the DE would help to identify the extent to which such differences are due to meeting order or the successful training of novice statisticians and data scientists to become effective interdisciplinary collaborators.

## 5. Discussion

This study adds to the literature by being the first evaluation of the effectiveness of the ASCCR framework for training statistics and data science collaborators. Our guiding research question for this study was, *In which ways and to what extent is the ASCCR framework useful for training novice statistics and data science students to become effective interdisciplinary collaborators?* Our video review of ten initial collaboration meetings provides evidence for the *ways* in which novices are effective. Novices performed well on projects by approaching collaborations with beneficial Attitudes, adequately Structuring meetings, thinking about both the qualitative and quantitative aspects of the Content of the project, using effective Communication skills, and by striving to create strong Relationships.

We believe that our results also provide evidence for the *extent* that novice students can be effective collaborators and answer our second research question, *During initial collaboration meetings, how well do novices perform compared to an expert collaborator?* Novices perform nearly as well as an expert collaborator in the initial meeting, with opportunities for improvement in the Communication and Content components of ASCCR. Over the course of a project, novices consistently perform well, generally out-scoring the expert on end-of-project feedback surveys.

A key finding from this study is that all 10 projects we investigated were successful. Previous research indicates that 94% of domain experts coming to LISA found the collaborations to be helpful and 93% were satisfied (Vance 2022a; b; c). So having all 10 projects deemed "successful" is not surprising. It does, however, limit our ability to identify components of ASCCR that—when not done well—lead to unsuccessful collaborations.

Based on the results of this study, we provide recommendations below for teaching interdisciplinary collaboration and running a statistics and data science consulting/collaboration laboratory or center (aka, a "stat lab"). If the collaboration course is not associated with a stat lab—and therefore does not have a stream of incoming projects for the students to work on—we recommend assigning pairs of students to find their own domain experts for one "midterm" project and one "final" project. Anyone with a genuine desire or need for statistics/data science help to answer a research question or make a data-driven decision can qualify as a domain expert and can provide an authentic collaboration experience for the students.

## 5.1. Recommendations for Teaching Collaboration

Our results provide evidence that even though the expert collaborator outperformed the novices initially, the novices closed the gap during subsequent meetings. This underscores how collaboration is a developmental process. Initial hiccups may be overcome by consistently applying ASCCR skills throughout the collaboration. This is good news for educators: effective interdisciplinary collaboration can be taught and learned. Below we briefly discuss recommendations for teaching the ASCCR framework to undergraduates and graduate students in a collaboration, consulting, or capstone course.

*Attitude.* We found that the rubric scores for the novices and the expert in 16 initial collaboration meetings were highest for the Attitude items (see Table 3). As was evident from the videos, a positive collaborative attitude can set the tone for a successful collaboration. In other words, effective attitudes can help lead a collaboration in a positive direction and can mask or offset weaknesses and mistakes in other areas. We recommend discussing with students which attitudes may be beneficial or detrimental for collaboration. Encourage your students to commit to being helpful to the domain expert and to generate a personal interest in the project. Focusing on what the domain expert wants/needs and how the student can help can ease the negative inner voice common with beginning students who lack self-efficacy in collaboration. Encourage your students to adopt a growth mindset (Dweck 2008). Asking the domain expert *why* they are interested in the project can create better shared understanding about the project and help the student get excited about the project, which in turn can help the student reframe any barriers or obstacles associated with the project into opportunities for learning and growth.

*Structure*. We recommend enabling students to practice applying the POWER structure during in-class role plays before doing it in real collaborations. In most of the videos, we observed that novices successfully transferred their Structure skills from the classroom to their real projects. Following the POWER structure and explicitly asking the domain expert what they wanted to accomplish during the initial meeting helped to focus the ensuing discussion. Setting aside time at the end of the meeting to summarize key decisions and next steps also focused the discussion in the initial meeting toward accomplishing concrete tasks such as understanding enough of the qualitative context of the problem ($Q_1$) to make headway on the quantitative ($Q_2$) elements of the problem.

*Content.* Most beginning students are really excited about the opportunity to apply their newly learned quantitative methods ($Q_2$) and often jump past understanding the problem context ($Q_1$) sufficiently. The novices on Project 2 illustrated this impetuosity:

> And then we can run through ridge regression, lasso regression and elastic net for all of them and we can compare them and see if there's anything notable between them. […] What are you actually looking for? What's your research question generally?

During the training of these specific novices, we required them to read a four-page summary of the $Q_1Q_2Q_3$ workflow (Vance 2019) and emphasized this workflow throughout our in-class discussions of projects. However, the biggest opportunity for improvement we observed in the novices' videos was not fully understanding $Q_1$ before thinking about and discussing $Q_2$. To address this in our *Statistical Collaboration* course, we now require students to read a recently published full-length paper about the $Q_1Q_2Q_3$ workflow (Vance et al. 2025), and we recommend this for others. We are also implementing the recommendations in that paper to teach the $Q_1Q_2Q_3$ workflow in precursory courses (e.g., Statistical Learning). As a result, in future semesters, many students in *Statistical Collaboration* will already have had practice thinking about how the qualitative context ($Q_1$) of a problem influences the use of the quantitative methods ($Q_2$) they are learning and the qualitative conclusions ($Q_3$) they can make.

*Communication*. Another area in which the expert collaborator outshone the novices was in Communication, specifically in verifying her understanding of the project with the domain expert and asking great questions. Yet, by the ends of the projects, the domain experts rated their novice collaborators highly on the feedback surveys. Our conclusions are that communication skills may take more time to develop than other skills in the ASCCR frame and that suboptimal Communication won't sink a project. In other words, if a collaborator has positive Attitudes,

Structures meetings well, focuses on the right Content, and has decent Communication skills, the project can be successful.

*Relationship.* Similarly, relationships with domain experts take time to develop. Our review of the initial meetings showed that the expert collaborator may have had a stronger relationship initially, but by the end of the project, the domain experts rated their relationship with the novices higher on average. We observed that—when done well—the Attitude, Structure, Content, and Communication components of ASCCR often lead to strong Relationships. We also observed the converse: starting projects with the intent to create a strong relationship with the domain expert can improve performance in the other components of a collaboration.

*Using the rubric with example quotes.* Although the video observation rubric was used in this study as an evaluation tool for effective collaboration, we recommend using it as an instructional tool. In Table A1 of Appendix 1, we have included student quotes as examples of scores of 4, 3, 2, or 1 for many of the rubric cells. A capstone or collaboration/consulting course could use this rubric (and ideally video clips of student collaborators) to help teach collaboration. Students could use the rubric to guide self-reflection on their own meeting videos. It could also be used as a guide for discussing instructional collaboration videos during class (e.g., videos from Sharp et al. (2021)). Instructors could show video clips representative of two different scoring levels of the same rubric element, ask students to score, and then facilitate a class discussion about what differentiated the levels of collaboration quality. Absent video clips, students could be asked to concoct plausible quotes that would score a 4.

## 5.2. Recommendations for Statistical Collaboration Laboratories

In this study, we observed evidence that novice collaborators—from undergraduate SDS majors to a variety of quantitatively-trained graduate students—can be effective interdisciplinary collaborators, even on their first few projects. Therefore, they can positively contribute to a stat lab, as described in Vance (2015). Below we briefly discuss implications of and recommendations from this study for stat labs, including the stat labs that follow our training methods in the LISA 2020 Global Network (Awe and Vance 2014; Awe et al. 2022; Vance et al. 2022c), which comprises 37 labs in 11 developing countries.

*Students can become effective collaborators almost immediately.* Notably, the novice (GU) pair on Project 9 had no statistical collaboration experience prior to meeting with the domain expert. Yet this pair outscored the expert collaborator on the Structure, Content, and Relationship components and tied the expert with perfect 4.0 rubric scores on Attitude and Communication. We recommend initial training and orientation in attitudes that promote collaboration, structuring collaboration meetings (including specifically focusing on/asking what the domain expert wants to accomplish), and appreciation of the importance of the qualitative ($Q_1$ and $Q_3$) components of projects to prepare novices. With such initial training, the novices will likely be a little worse than an expert collaborator, but will still provide value to the domain expert who ends the project satisfied with the collaborators and the stat lab. As we saw with Project 9, novices have the potential to surpass adequacy and achieve immediate excellence.

*Thoughtful pairing of novices on a project is important.* Our study involved novices with diverse backgrounds. A factor in our pairings was a sense by the course instructor of how the novices' technical and collaboration skills would complement each other's. In the videos, we observed several instances in which one novice was at a loss for words, but their partner stepped in to steady the situation. This reinforces our recommended practice of assigning two student collaborators to work together on most projects because one student can mitigate their partner's weaknesses. This thoughtful pairing of novices on projects may be why the rubric scores (see

Section 4) did not reveal any systematic differences in the performance of novice undergraduate and graduate students nor in their pairings (U2, GU, or G2).

*Prepare the domain expert for the initial meeting*. Another implication from this research is the need to prepare students to communicate with domain experts about the initial project meeting. DE3 discussed how the initial meeting with the novices helped her prepare for the initial meeting with the expert collaborator. She was more equipped to talk about her project and anticipated some of the questions the collaborator might ask after having a first experience with the novices. We recommend instructors work with their students to create a template or handout for domain experts to help them prepare for the initial meeting. This might include examples of questions the collaborators intend to ask during the initial meeting. Such a tool can help both the collaborators and the domain expert prepare for the initial meeting.

## 5.3 Limitations and Future Work

Our study was intentionally designed so that the expert collaborator had the advantage of meeting with the domain expert second, one or more days after the novices conducted the first initial meeting. On the other hand, the novices had the advantages of working with a partner and working with the domain expert for additional meetings. Research to determine the extent to which order of collaboration meetings might influence differences in rubric scores and how well novices perform alone versus as a two-person collaboration team could be helpful.

Although our sample of 10 collaboration projects was larger than the one project considered in Alzen et al. (2023), it is still a sample limited in size and the extent of the collaboration meetings under consideration. We only scored the first meeting for all of our collaboration projects. Additionally, it is possible that our sample of domain experts who self-selected to be a part of our study was biased in a way that caused their feedback survey scores to favor student collaborators, though nothing indicating such a bias was observed during video review. More likely, the domain expert feedback scores were biased in favor of the novices because they spent more time with the domain experts. More in-depth work that uses video data for the duration of the project would provide better indications about the ways and the extent to which students can be effective statistical collaborators.

We compared an expert with novice collaborators, all of whom had been trained in ASCCR. Ideally, to evaluate the effectiveness of the ASCCR framework, one might compare three groups: an expert to provide a standard for collaboration, novices who were not trained in ASCCR, and novices who were trained in ASCCR. Such comparisons would likely be fraught with difficulties, however.

Finally, our study sample contained no failed projects or bad meetings. This limits our ability to scrutinize the ways in which collaborators may be unintentionally sabotaging meetings and causing projects to fail. Future work might consider scrutinizing all mediocre or poor projects in a stat lab to attempt to quantify which components of collaboration are most likely to cause breakdowns in the project, as well as how to best repair any breakdowns (Zahn 2019). To get a larger sample of projects that compare an expert collaborator with novice collaborators, or to evaluate how well novices perform on real projects in different contexts, we encourage more educators to teach the ASCCR framework and for more stat labs to collaborate with each other. A multi-site study evaluating methods of teaching statistical collaboration would bolster evidence about the most effective approaches for educating and training future statistics and data science collaborators.

## 6. Conclusion

We compared the collaboration skills of novice students trained in the ASCCR framework against an expert collaborator, using a mixed-methods approach that included video analysis of initial meetings and end-of-project domain expert feedback. Our findings provide evidence that the ASCCR framework is an effective method for statistics and data science educators and students to develop essential interdisciplinary collaboration skills quickly. Our results reveal that novices can rapidly develop collaboration proficiency, performing comparably to an expert in the Attitude, Structure, and Relationships components, though experience seems to yield more nuanced skills in Content and Communication.

Crucially, our study demonstrates that initial imperfections do not preclude long-term success. Despite scoring lower than the expert collaborator on average for the initial meeting, the novices closed the gap, ultimately earning higher overall satisfaction scores from the domain experts by the conclusion of the projects. These findings should reassure both students and instructors that effective collaboration is a set of learned skills that can improve over time. We therefore urge statistics and data science educators to incorporate the ASCCR framework into their collaboration, consulting, or capstone courses, applying the practical curriculum recommendations provided in this paper. To build upon this evidence base, future research should focus on a multi-site study to provide evidence about the most effective approaches for educating and training future statistics and data science collaborators.

## Acknowledgements

We thank the domain experts and students of the Laboratory for Interdisciplinary Statistical Analysis (LISA) for their willingness to participate in this study. We thank Isabel Maloney for providing useful R code. We also thank the anonymous reviewers and the editors who provided helpful feedback.

## Funding

This material is based upon work supported by the National Science Foundation under Grant No. 1955109, Grant No. 2022138, and Grant No. 2044384 for the projects, “IGE: Transforming the Education and Training of Interdisciplinary Data Scientists (TETRDIS)”, “NRT-HDR: Integrated Data Science (Int dS): Teams for Advancing Bioscience Discovery”, and “CODE:SWITCH: Integrating Content and Skills from the Humanities.” This work was also partially supported by the United States Agency for International Development under Cooperative Agreement #7200AA18CA00022 for the project, “LISA 2020: Creating Institutional Statistical Analysis and Data Science Capacity to Transform Evidence to Action.”

## Data Availability Statement

The rubric, example checklist, domain expert survey, deidentified quantitative data, and code that support the findings of this study are openly available on the Open Science Framework at https://osf.io/ajc7g. Video transcripts are unavailable due to IRB restrictions.

## Ethics Approval

This study was approved by the Institutional Review Board of the University of Colorado Boulder (Protocol 18-0554). Informed consent was obtained from all participants (including the novice students, statistical expert, and domain experts) prior to data collection. The instructor was blinded to the students’ and domain experts’ consent to participate in the study.

## Generative Artificial Intelligence (AI) Use

The authors acknowledge the use of Google's Gemini 3.5 during the revision of this manuscript. Specifically, the tool was utilized to wordsmith, refine language, and improve the clarity and flow of the paper's title, abstract, and conclusion. The authors have rigorously reviewed and edited all AI-generated outputs and take full responsibility for the originality, validity, and accuracy of the final text.

## Disclosure Statement

The authors report there are no competing interests to declare.

# A1. ASCCR Observation Rubric

**Table A1.** ASCCR Observation Rubric with example quotes from this study's projects.

| **Attitudes** | | | | |
|---|---|---|---|---|
| | **4**<br>**Mastery** | **3**<br>**Competent** | **2**<br>**Developing** | **1**<br>**Minimal** |
| Collaborative Attitude | A collaborative attitude is evident throughout the meeting (e.g. Focus is on shared goals and meeting the domain expert's [DEs] goals.)<br><br>*Project 1:*<br>*Yeah, those notes [that the DE provided] are nice. I was initially kind of taking notes on the meeting notes page [created by the collaborators], but honestly the notes you [the DE] gave us are pretty solid.* | Collaboration is clearly a priority, but at times the LISA collaborators forget to ask the DE for their input. | Collaboration is evident in some elements of the meeting, but collaborators often fail to invite the DE to contribute. | Collaborators tell the DE what will happen in the meeting. DE is not included in the goals for the meeting or meeting activities. |
| Self-Attitude | Collaborators present selves as experts in their | Collaborators generally present | Collaborators present selves as struggling | Collaborators present selves as struggling |
| | own fields as well as individuals who can learn about a new field from a DE.<br><br>*Project 7:*<br>*I've never worked with a shape file. I guess could you give some kind of background on that, because I'm mainly an R Studio person, so just curious on the shape file, what it's like and what system you even use for that.* | confidence in skills. They occasionally make self-deprecating comments, but this is not a dominant attitude in the meeting.<br><br>*Project 4:*<br>*I think it would be doing yourself a disservice to throw out a large part of your data […].But there's a technique called censoring, which does allow for a bit of handling for stuff that is longer than you can measure.* | with effective statistical collaboration from EITHER a technical OR interdisciplinary collaboration standpoint. | with completing effective statistical collaboration from both a technical and interdisciplinary collaboration standpoint. |
| Attitude toward Domain Expert | DE is treated as an equal contributor or even greater expert than the collaborators in the meeting.<br><br>*Project 2:*<br>*It really looks like you have an awesome start there as well. […] We'll make sure that everything's been appropriate, that you've done, we're not dealing with a violation of some other assumption along the way you might've missed.* | Collaborators occasionally consider DE expertise. | Collaborators rarely acknowledge DE expertise. | Collaborators present themselves as the only experts in the room. Little effort is made to leverage the expertise of the DE. |

| Structure | | | | |
|---|---|---|---|---|
| Preparation | Collaborators show that they have prepared for the meeting by, for example, stating the problem in their own words and providing interest in the topic as an indication of having thought about the problem prior to the meeting. | Some preparation efforts are evident. The collaborators speak fluidly about the project premise but do not make additional connections between the project and their prior work.<br><br>*Project 7:*<br>*I guess when you're talking about forecasting, are you talking about in a sense your linear regression models that you had mentioned on your document [that you sent ahead of this meeting] of predicting for that or are you wanting, I guess what do you want from forecasting? What are you meaning by that?* | Minimal preparation efforts are evident. For example, collaborators may read from the project request form. | No evidence of preparing for the meeting evident. Collaborators appear to have no prior knowledge of project. Collaborators are ill-prepared to engage in the project. |
| Open | Meeting is opened with a friendly tone. Time is spent discussing everyone's wants for the meeting.<br><br>*Project 1:* | Meeting is opened with a friendly tone. Time is spent discussing everyone's wants for the meeting, but may be hurried, rushed, or incomplete. | Collaborators may or may not take time for friendly interaction at the beginning of the meeting. If goals/an agenda are shared, the collaborators tell the DE the goals rather | Collaborators don't take time for friendly interaction at the beginning of the meeting. Goals and/or an agenda are not shared at the beginning of the meeting. |
| | *And so our overall goals for this meeting were to kind of get a better understanding of your research and kind of the statistical questions involved with it. I was also wondering if you had any specific goals that you wanted to accomplish this meeting?* | | than including them in the process.<br><br>*Project 1:*<br>*DE: I should learn which of you is [Collaborator 1] and who is [Collaborator 2]?*<br>*Collaborator 1: Yeah, I guess we could introduce ourselves.*<br>*Collaborators 2: Yeah, that would be nice.* | |
| Work | Collaborators shape the bulk of the meeting in a way that specifically responds to the wants agreed upon during the opening conversation. If all agreed upon wants are not addressed, the collaborator makes a plan for addressing them at a later time. | The collaborators shape the meeting to respond to the DEs wants agreed upon during the opening conversation, but if some go unmet, no plan is made to address them in the future. | Some effort is made to address DE wants during the meeting, but at times the collaborators jump to statistics before answering all DE requests/questions. | Collaborators ignore DE wants and jump into solving the statistics for the DE. Details of $Q_1$ are ignored. |
| End | 10–20% of meeting time is reserved for the closing conversation. All applicable checklist elements of “end” are present. | Time is reserved for the closing conversation, but it may be insufficient. Some checklist elements of “end” are missed.<br><br>*Project 6:* | Collaborators attempt to cover some checklist elements of ending a meeting via the POWER structure, but time was not reserved. The conversation is rushed and incomplete. | No time is reserved for meeting closure activities. Collaborators do not attempt to cover checklist elements of ending a meeting via POWER structure. |
| | | *Okay. We also do only have about five minutes left in this meeting, and I wanted to kind of just take a minute to summarize what we had gone over and make sure we covered at least a little bit of what you wanted to and then see if we want to kind of schedule another meeting for after spring break. I mean, yeah, thank you so much for giving us this background, for showing the data, so it was very helpful. I think we have a better grasp on this now.* | | |
| Reflection | Collaborators allow space for and invite reflection regarding success of meeting. | Collaborators ask for reflections on effectiveness of meeting but do not provide sufficient time for the conversation. | Collaborators ask for reflections on effectiveness of meeting but do not reserve any time for the conversation. | Collaborators do not provide space for meeting reflection with the DE. |
| POWER Structure | Collaborators run the meeting in ways that lessen the cognitive load for the participants. All applicable elements of POWER structure are present. | Collaborators run the meeting in ways that lessen the cognitive load for the participants. All sections of the POWER structure are | Collaborators attempt to address every section of the POWER structure, but few checklist elements of each phase are present. | Collaborators do little to nothing to appropriately structure the meeting for efficiency and effectiveness. Very few, if any, checklist |
| | | present, but there are several checklist elements missing across the sections. | | elements of the POWER structure are evident. |
| **Content** | | | | |
| Q1 | The collaborators paraphrase/state all of the appropriate elements below: (1) the problem, (2) how the solution to the problem will be used, (3) ideal data, (4) actual data, (5) the reasons the DE finds the problem important and/or interesting, (6) a reason the collaborators also find the problem interesting or important, and (7) which types of statistical analysis would be most useful.<br><br>*Project 4:*<br>*Let's get a little more context in mind before we jump into the data, if that's all right? […] If you could tell us more about what your project is and sort of the nature of the experiments, I think that would be helpful.* | The collaborators address all elements of Q1 appropriate for the meeting, but there may be some lack of clarity on either the part of the collaborators or the DE. | Collaborators address more than half of the appropriate elements of Q1 before moving on to Q2. Clarity for both the DE and collaborators may or may not be present.<br><br>*Project 1:*<br>*Bayesian approach could work for sure. It's hard to say without seeing more of the data and what's going on. But yeah, Bayesian is, yeah, like you said, where you have a prior expectation and put in some new data and see how it shifts for your posterior distribution. But yeah, it really depends on what kind of assumptions you want to make for the prior.* | Collaborators address fewer than half of the elements of Q1 that were appropriate for the meeting and move on to Q2 before gaining any real understanding of Q1. Clarity for both the DE and collaborator may or may not be present.<br><br>*Project 2:*<br>*And then we can run through ridge regression, lasso regression and elastic net for all of them and we can compare them and see if there's anything notable between them. […] What are you actually looking for? What's your research question generally?* |
| Q2 (if applicable) | The collaborators suggest an appropriate statistical technique and appropriately execute it to address the DEs wants. | The collaborators suggest an appropriate statistical technique but does not execute it without error. | The collaborators suggest an inappropriate statistical technique but executes it appropriately. | The collaborators suggests an inappropriate statistical technique and execute it with errors. |
| Q3 (if applicable) | Collaborators translates how the statistical analysis answers the research question qualitatively in a way that the DE understands. Collaborators check DE understanding in a respectful manner. | Collaborators translate how the statistical analysis answers the research question qualitatively in a way that the DE may understand. The collaborators check with the DE for understanding, but it is unclear if the DE does, in fact, understand. | Collaborators provide statistical results, but translation is ineffective. An attempt is made to connect results to the research questions, but the connection is unclear. Collaborators do not check to see if the DE understands the results or checks in a way that does not honor the DE. | Collaborators provide statistical results but do not translate them or connect them to the research questions. There is no check for DE understanding. |
| **Communication** | | | | |
| Questioning | Collaborators effectively use questioning throughout the meeting to provide clarity for both themselves and the DE. Collaborators ask “great” questions that both elicit useful information and strengthen the relationship. | Questioning is attempted frequently during the meeting but is sometimes insufficient for gaining clarity for collaborators or DE.<br><br>*Project 4:*<br>*What are these experiments that you have been doing with them? How are you generating the data* | Collaborators ask questions to help clarify for selves or the DE but not both regularly throughout the meeting. | Questioning is not evident in the meeting. |
| | | *and what question are you trying to answer?* | | |
| Listening | The collaborators listen attentively throughout the meeting. When the collaborators speak, it is in direct response to the previous comment by the DE. | The collaborators listen attentively throughout the meeting. When the collaborators speak, it is in direct response to the previous comment by the DE. However, sometimes the collaborators do not fully respond to DE comments. | The collaborators occasionally jump ahead of the DE to talk about their own desired topics. Responses sometimes do not make sense after DE comments. | The collaborators regularly jump ahead of the DE to talk about their own desired topics. Responses do not make sense after DE comments. |
| Paraphrasing | Collaborators paraphrase regularly throughout the meeting to clarify both their own language and the language of the DE. | Collaborators paraphrase regularly throughout the meeting for clarity but may only do this for the DE or themselves.<br><br>*Project 4:*<br>*So then just to test my understanding, you're looking at the time they spend in the closed arms of the plus maze as one measure of anxiety, you're looking at the time to reach the shelter and the time to leave the shelter?* | Collaborators occasionally paraphrase during the meeting to add clarity for either themselves or the DE. | Collaborators rarely, if ever, paraphrase during the meeting. |
| Summarizing | Collaborators summarize regularly throughout the meeting to solidify discussion topics and decisions made. | Collaborators summarize regularly throughout the meeting but may miss some key moments that would have benefitted from summarizing. | Collaborators occasionally summarize during the meeting but individuals will leave the meeting with lack of clear understanding of the discussions and decisions. | Collaborators rarely, if ever, summarize during the meeting. |
| Explanation | Collaborators use analogies, diagrams, examples, plain language, and technical definitions (ADEPT) effectively to explain statistical terms to non-statisticians, as appropriate. | Collaborators attempt many elements of ADEPT but some may not be effective in translating the statistics well. | Collaborators attempt some elements of ADEPT but some may not be effective in translating the statistics well. | Collaborators do not use any elements of ADEPT to explain statistics to DE. |
| | **Relationship** | | | |
| Relationship | Collaborators and DE both make positive comments during the meeting about the relationship.<br><br>*Project 7:*<br>*DE: I'm so grateful for you guys, and thank you so much ahead of time for your patience and your just willingness to explore. I just, I'm so just up for whatever, and it will benefit this project regardless, and I'm just grateful for your help* | Collaborators or DE make positive comments during the meeting about the relationship.<br><br>*Project 4:*<br>*I really appreciate y'all's help with this. It's been super helpful and I'm excited to tackle this now, so I appreciate y'all.* | Positive comments are made during the meeting, but it is not clear if they refer to the collaboration specifically. | No comments are made during the meeting about the relationship between the collaborator and DE. |
| | *because I did not feel confident about this at all, and now I feel so much better.*<br>*[...]*<br>*Collaborator: We're really excited on this project too. It's really interesting.* | | | |

## A2. Domain Expert Feedback Survey

Q1 Approved IRB informed consent text here

- ○ Agree. Take me to the survey. (1)
- ○ Do not agree. Leave survey. (2)

Q3 You are being asked to take this survey because you worked with LISA on {project} during {semester}. You worked with the following statistical collaborator(s): {statistical collaborator 1}; {statistical collaborator 2}. Please answer all questions with your experience on this project with these collaborators in mind.

Q4 Rate how much you agree with the following statements:

| | Strongly disagree (1) | Disagree (2) | Somewhat disagree (3) | Somewhat agree (4) | Agree (5) | Strongly agree (6) |
|---|---|---|---|---|---|---|
| The service I received from LISA was helpful for my project (1) | ○ | ○ | ○ | ○ | ○ | ○ |
| I am satisfied with my overall LISA experience (2) | ○ | ○ | ○ | ○ | ○ | ○ |

Q5 Does the help you received warrant:

| | Yes (1) | Maybe (2) | No (3) |
|---|---|---|---|
| Acknowledgement of your collaborator in a thesis or dissertation (1) | ○ | ○ | ○ |
| Co-authorship on a paper with your collaborator (2) | ○ | ○ | ○ |
| Acknowledgement of your collaborator in a paper (3) | ○ | ○ | ○ |
| A monetary contribution from an existing grant (4) | ○ | ○ | ○ |
| Potentially a co-authorship on a paper with your collaborator in the future (5) | ○ | ○ | ○ |

Q14 Please explain any items for which you selected "Maybe" in the question above

________________________________________________________________

Q15 How useful were your meetings with LISA collaborators?

- ○ Not at all (1)
- ○ A little (2)
- ○ Somewhat (3)
- ○ Moderately (4)
- ○ Very (5)
- ○ Extremely (6)

Q17 On a scale from 0 to 10, how likely are you to recommend LISA to a friend or colleague

- ○ Not at all Likely (0) (0)
- ○ 1 (1)
- ○ 2 (2)
- ○ 3 (3)
- ○ 4 (4)
- ○ 5 (5)
- ○ 6 (6)
- ○ 7 (7)
- ○ 8 (8)
- ○ 9 (9)
- ○ Extremely Likely (10) (10)

Q19 Rate how much you agree with the following statements

| | Strongly disagree (1) | Disagree (2) | Somewhat disagree (3) | Somewhat agree (4) | Agree (5) | Strongly agree (6) |
|---|---|---|---|---|---|---|
| I felt welcomed during meetings with LISA collaborators (1) | ○ | ○ | ○ | ○ | ○ | ○ |
| My interactions with LISA collaborators outside of meetings were positive (2) | ○ | ○ | ○ | ○ | ○ | ○ |

Q6 Thinking about your own understanding of {your project} before and after meeting with LISA, please rate

| | Very Low (1) | Moderately Low (2) | Low (3) | High (4) | Moderately High (5) | Very High (6) |
|---|---|---|---|---|---|---|
| Your understanding of your research BEFORE meeting with LISA | ○ | ○ | ○ | ○ | ○ | ○ |
| Your understanding of your research AFTER meeting with LISA | ○ | ○ | ○ | ○ | ○ | ○ |
| Your understanding of statistical aspects of your research BEFORE meeting with LISA | ○ | ○ | ○ | ○ | ○ | ○ |
| Your understanding of statistical aspects of your research AFTER meeting with LISA | ○ | ○ | ○ | ○ | ○ | ○ |

Q9 You worked with {statistical collaborator 1} and {statistical collaborator 2} on this project. Thinking about your collaborator(s) understanding of {your project} before and after meeting, please rate

| | Very Low (1) | Moderately Low (2) | Low (3) | High (4) | Moderately High (5) | Very High (6) |
|---|---|---|---|---|---|---|
| Your collaborators' general understanding of your research area BEFORE meeting | ○ | ○ | ○ | ○ | ○ | ○ |
| Your collaborators' understanding of your research AFTER meeting | ○ | ○ | ○ | ○ | ○ | ○ |
| Your collaborators' general understanding of the statistical aspects of your research area BEFORE meeting | ○ | ○ | ○ | ○ | ○ | ○ |
| Your collaborators' understanding of the statistical aspects of your research AFTER meeting | ○ | ○ | ○ | ○ | ○ | ○ |

Q20 How would you rate the collaborators' overall contribution to the project?

○ Very Low (1)

○ Moderately Low (2)

○ Low (3)

○ High (4)

○ Moderately High (5)

○ Very High (6)

Q22 How would you rate the strength of your relationship with the collaborators?

- ◯ Very Low (1)
- ◯ Moderately Low (2)
- ◯ Low (3)
- ◯ High (4)
- ◯ Moderately High (5)
- ◯ Very High (6)

Q8 Please give specific feedback on these collaborators. What did they do well? What were their strengths? What could they do to improve?

________________________________________________________________

Q18 To what extent did your work with LISA advance your research or data-driven decision making? How are you thinking about your research and where you might go next after working with LISA?

________________________________________________________________

Q10 Each year, LISA gives an Outstanding LISA Collaborator award. This award will recognize a student for his or her outstanding work on a project or projects involving LISA. If you would like to name any of the collaborators you worked with on this project, please do so below and explain why you nominated that student.

________________________________________________________________

Q12 Do you have any additional comments or general feedback about LISA? Is there anything else the Director of LISA should know?

________________________________________________________________

----End of Domain Expert Feedback Survey---